\documentstyle[12pt]{article}

\def\bc{\begin{center}}
\def\ec{\end{center}}
\def\be{\begin{equation}}
\def\ee{\end{equation}}
\def\myappendix{\par
 \setcounter{section}{0}
 \setcounter{subsection}{0}
 \setcounter{equation}{0}
 \setcounter{table}{0}
 \def\appendixname{Appendix}
 \def\appesection{\setcounter{equation}{0}\section}
 \def\@thesection{\Alph{section}}
 \def\thesection{\appendixname\hskip 1.10ex\Alph{section}}
 \def\thesubsection{\@thesection.\arabic{subsection}}
 \def\theequation{\@thesection.\arabic{equation}}
 \def\thetable{\@thesection.\arabic{table}}}

\newcommand{\labar}{\overline{\Lambda}}

\newcommand{\energy}{\bar{h}(x)\, D_{4}\, h(x)}
\newcommand{\unity}{\bar{h}(x)\, h(x)}

\newcommand{\beq}{\begin{equation}}
\newcommand{\eeq}{\end{equation}}
\newcommand{\beqn}{\begin{eqnarray}}
\newcommand{\eeqn}{\end{eqnarray}}

\def\vdir{v\kern-7.8pt\Big{/}}
\def\pdir{p\kern-7.8pt\Big{/}}

\begin{document}
\pagestyle{empty} 
\vspace{-0.6in}
\begin{flushright}
CERN-TH/96-175\\
ROME prep. 96/1145 \\
SHEP 96-15\\ 
 FTUV 96/22 - IFIC 96/24
\end{flushright}
\vskip 0.2 cm \centerline{\LARGE{\bf{A High-Statistics Lattice
      Calculation of}}} \vskip 0.2cm
\centerline{\LARGE{\bf{$\lambda_1$ and $\lambda_2$ in the $B$ meson}}}
\vskip 0.2cm 
\centerline{\bf{V. Gim\'enez$^{1}$, G.  Martinelli$^{2}$ and
C. T. Sachrajda$^{3,*}$}} 
\centerline{$^1$ Dep. de Fisica Teorica and IFIC, Univ. de Valencia,} 
\centerline{Dr.  Moliner 50,
  E-46100, Burjassot, Valencia, Spain.} \centerline{$^2$ Dip. di
  Fisica, Univ. ``La Sapienza" and} \centerline{INFN, Sezione di Roma,
  P.le A. Moro, I-00185 Rome, Italy.} \centerline{$^3$ Theory
  Division, CERN, 1211 Geneva 23, Switzerland.} 
\abstract{ We present a high-statistics lattice calculation of the
  kinetic energy $-\lambda_1/2 m_b$ of the heavy quark inside the
  $B$-meson and of the chromo-magnetic term $\lambda_2$, related to
  the $B^*$--$B$ mass splitting, performed in the HQET.  Our results
  have been obtained from a numerical simulation based on 600 gauge
  field configurations generated at $\beta=6.0$, on a lattice volume
  $24^3 \times 40$ and using, for the meson correlators, the results
  obtained with the SW-Clover $O(a)$ improved lattice action for the
  light quarks.  For the kinetic energy we found $-\lambda_1=\langle B
  \vert \bar h (i\vec{D})^{2} h \vert B \rangle /(2 M_B )=-(0.09 \pm
  0.14)$~GeV$^2$, which is interesting for phenomenological 
  applications.  We also find $\lambda_2= 0.07 \pm 0.01$ GeV$^2$,
  corresponding to $M^2_{B^*}-M^2_B= 4 \lambda_2= 0.280 \pm 0.060 $
  GeV$^2$, which is about one half of the experimental value. The
  origin of the discrepancy with the experimental number needs to be
  clarified.}
\vskip 0.5cm 
\centerline{$^*$ On leave of absence from
  Dept. of Physics, University of Southampton, Southampton SO17 1BJ,
  UK.} 
\vfill\eject 
\pagestyle{empty}
\clearpage 
\setcounter{page}{1}
\pagestyle{plain} 
\newpage \pagestyle{plain} \setcounter{page}{1}
 
\section{Introduction} \label{intro}

In a previous study \cite{cgms} we presented several results for the
$B$-meson binding and kinetic energies, obtained by using the lattice
version of the Heavy Quark Effective Theory (HQET) \cite{eihil}.  The
calculation required the non-perturbative subtraction of the power
divergences present in the matrix elements of the Lagrangian operator
$\bar h D_4 h$ and of the kinetic energy operator $\bar h \vec D^2 h$,
following the strategy outlined in ref.~\cite{ms}.  Good results were
found for the binding energy $\labar$, which have been confirmed and
improved by the higher statistics analysis of ref.~\cite{gms}. On the
other hand, given the low statistical sample at our disposal in
ref.~\cite{cgms}, the determination of $\lambda_1$ was quite poor and
resulted only in a weak upper bound of $\vert \lambda_1 \vert \le 1$
GeV$^2$.

\par In this paper, we present a high-statistics calculation
of both $\lambda_1$ and $\lambda_2$, based on a sample of $600$
configurations at $\beta=6.0$, on a lattice volume $24^3 \times 40$.
Using the same renormalized kinetic energy operator as in dimensional
regularization (e.g. as in the $\overline{MS}$ scheme)~\footnote{ Our
  kinetic energy operator is made finite by subtracting
  non-perturbatively the power divergences, and then by matching it to
  the $\overline{MS}$ operator using perturbation theory.} we find
$\lambda_1 = 0.09\pm 0.14$~GeV$^2$.  Even though the determination of
$\lambda_1$ is still not very precise, it is however interesting for
phenomenological applications, and for comparison with other
theoretical predictions.
Our result for the matrix element of the chromo-magnetic operator,
$\lambda_2 \sim 0.07$ GeV$^2$, obtained with  rather large
statistics, is in excellent agreement with the latest result of the
UKQCD collaboration \cite{ukqcdl2}, which had been performed with a
smaller statistical sample, but at a smaller value of the lattice
spacing~\footnote{The original version of the UKQCD preprint
  \cite{ukqcdl2}, had an error of a factor of 2 in the value for
  $\lambda_2$. The corrected value of the mass splitting
is  $M_{B^*}^2-M_B^2 = 0.281\pm 0.015\pm
  0.040$~GeV$^2$.}.  A value of $\lambda_2 \sim 0.07$~GeV$^2$ corresponds to
$M_{B^*}^2-M_B^2 \sim 0.280$ GeV$^2$, which is about one half of  the
experimental number, $0.485$~GeV$^2$.  In our opinion there are three
possible sources which may contribute to the explanation of the
discrepancy. The first is the fact that, at one-loop order, the factor
necessary to match the chromo-magnetic lattice operator of the HQET to
its continuum counter part is very large \cite{fh}, and it is possible
that the higher-order terms modify this factor significantly. A second
source of systematic error is quenching.  The final possibility is
that the discrepancy is due to a physical reason, i.e. that
$M_{B^*}^2-M_B^2$ receives significant contributions from higher-order
terms in the HQET. It seems to us unlikely that any of the three
sources alone could explain a discrepancy of about a factor of two
between the lattice and experimental values of $\lambda_2$.

\par The calculation of $\lambda_1$ and $\lambda_2$ proceeds in two different
ways. The chromo-magnetic operator that enters in the calculation of
$\lambda_2$ is logarithmically divergent in the ultra-violet cut-off,
i.e. the inverse lattice spacing $1/a$. Its multiplicative
renormalization constant can in principle be computed in perturbation
theory \cite{fh}, or with the non-perturbative technique proposed in
ref.~\cite{NP}. In the calculation of $\lambda_1$, instead, one has to
make a non-perturbative subtraction, related to the presence of a
quadratic divergence in the matrix elements of $\bar h \vec D^2 h$
\footnote{ In order to define a finite operator $\bar h \vec D^2 h$
  the subtraction of a linearly divergent term is also needed. This
  subtraction, however, does not enter in the calculation of
  $\lambda_1$, see below.}.  The subtracted operator only requires a
finite multiplicative renormalization constant, which is different
from unity because of the breaking of the reparametrization invariance
of the HQET on the lattice \cite{mms}--\cite{ms2}.

\par Although it may be convenient to introduce the parameter $\lambda_1$
(or the binding energy $\labar$), it is not necessary to do so.  One
can either relate two or more physical quantities directly to the
required precision in the heavy-quark expansion, or determine physical
quantities directly from some non-perturbative computation, such as
for example from the calculation of the bare lattice matrix elements
in numerical simulations. Indeed the actual values of $\labar$ and
$\lambda_1$ by themselves have no physical meaning, since they depend
on the definition that one decides to adopt for the corresponding
operators, and do not give us any relevant physical information.
Predictions for physical quantities can be obtained only by combining
these parameters with the coefficients and matrix elements of the
dominant operators.  This is to be contrasted with $\alpha_s$ or
$\lambda_2$, which correspond directly to measurable physical
quantities up to higher power corrections in the inverse mass of the
heavy quark. The introduction of a consistent definition of
$\lambda_1$, which is independent of the regularization and is of
$O(\Lambda_{QCD}^2)$, presents us, however, with an intuitive estimate
of the size of the corrections and can be used for comparison with
other definitions currently used in heavy-flavour physics.

\par The plan of the paper is the following. In sec. \ref{l1def} we
recall the relevant formulae, which define the non-perturbative
procedure for renormalizing the operators $\bar h \vec D^2 h$
\cite{ms} and the method used to compute the hadronic matrix element
of the subtracted operator in order to obtain $\lambda_1$; in the same
section we also describe the method used to obtain $\lambda_2$; in
secs. \ref{numerical1} and \ref{numerical2} we present the numerical
results for $\lambda_1$ and $\lambda_2$, respectively, together with a
comparison with other calculations of the same quantities.  Finally in 
section \ref{conclu} we present our conclusions.

\section{Definition of the renormalized $\lambda_1$ and $\lambda_2$} 
\label{l1def}
In this section we define the prescriptions that we 
will use to calculate $\lambda_1$ and $\lambda_2$. We also
recall the procedure used to extract the hadronic matrix elements 
of the renormalized operators from suitable two- and three-point
correlation functions.

\par The  renormalization of $\bar h \vec D^2 h$
can be implemented by imposing suitable non-per\-tur\-ba\-tive
renormalization conditions on its quark matrix elements in the Landau
gauge~\cite{cgms,ms}.  This is not necessary and it would certainly be
possible to obtain physical predictions using the matrix elements of
the bare lattice operator, which diverges as powers of $1/a$.  Indeed,
in a theory regulated by a dimensionful cut-off, it is consistent not
to perform the subtractions of the power-divergent terms at all, but
to work with the bare operators and to compute the coefficient
functions (which will therefore contain powers of the cut-off) in
perturbation theory \cite{ms,bigi}. In this case, however, the matrix
elements in the effective theory are divergent in the ultra-violet
cut-off and depend on the regularization. Therefore they cannot be
interpreted as ``phy\-si\-cal" quantities, in contrast with the approach
that we have adopted in ref.~\cite{cgms} and here, and with the use of
$\lambda_1$ that is made in the literature. Since $\lambda_1$ enters
in the theoretical predictions for several quantities 
relevant to the physics of hadrons containing the $b$-quark (hadron
spectroscopy, inclusive decays, form factors in exclusive weak and
radiative decays, etc.), we find it convenient to introduce a quantity
that is finite and independent of the regularization \cite{cgms,ms}.

\par In order to remove the power divergences from the kinetic energy
operator, we have imposed on the relevant operator a renormalization
condition that corresponds to the ``physical" requirement $\langle
h(\vec p=0)\vert \bar h \vec D_s^2 h \vert h(\vec p=0) \rangle = 0$,
where $\bar h \vec D_s^2 h$ is the subtracted kinetic energy operator
\cite{ms}.  Since a complete discussion of the main theoretical
aspects of the proposed non-perturbative renormalization conditions
and matching of the operators has been presented in our previous
papers on this subject \cite{cgms,ms}, only those formulae which are
necessary to the understanding of the numerical results are reported
in the following.

\par The subtracted  kinetic operator 
$\bar h \vec D_s^2 h$, which is free of power divergences, has the
form
\begin{eqnarray}
\bar{h}(x)\, \vec{D}_s^{2}\, h(x)  &=& \bar{h}(x)\, \vec{D}_x^{2}\, h(x)
- \frac{c_1}{a}\, \left[\frac{1}{(1+a \, \delta m)}
\Bigl(\, \energy\, +\, \delta m\, \unity\, \Bigr)\right]\, 
\nonumber \\ &-&\, \frac{c_{2}}{a^{2}}\, \unity, 
\label{eq:d2ren}
\end{eqnarray}
where for the discretized version of $D_4$ and $\vec{D}_x^{2}$ we have
taken 
\be \left[ D_4\right]_{\alpha \beta}=\frac{1}{a} \left( \delta_{\alpha
    \beta} \, \delta_{x,x}- U^{4 \, \dag}_{\alpha \beta}(x-a \hat t )
  \delta_{x, x\, -\, a\,\hat{t}} \right) \, ,\ee
and
\begin{eqnarray}
\left[\vec{D}_x^{2}\right]_{\alpha \beta} &=& \frac{1}{a^{2}}\, 
\sum^{3}_{k=1}\, \left(\, U^{k}_{\alpha \beta}(x)\, 
\delta_{x, x\, +\, a\,\hat{k}}
\,+\, U^{k\, \dag}_{\alpha \beta}(x\, -\, a\, \hat{k})\, 
 \delta_{x, x\,
 -\, a\,\hat{k} }\,-\, 2\, \delta_{\alpha \beta} \, \delta_{x, x}\right) .
\end{eqnarray}
The term in square parenthesis in eq.~(\ref{eq:d2ren}) 
is the subtracted lattice Lagrangian operator 
at lowest order in the HQET and $\delta m $ is the suitable
mass counter-term needed to eliminate the linear divergence
present in the unsubtracted Lagrangian \footnote{ For simplicity
we give here the expression corresponding to the unimproved lattice
Lagrangian for the heavy quark; the more complicated expression
necessary in the improved case can be found in ref.~\cite{ms}.
In the following, the numerical results 
will be given for the improved case only.}.
\par In eq.~(\ref{eq:d2ren})
the constants $c_{1}$ and $c_{2}$ are functions of the bare
lattice coupling constant $g_0(a)$. They have been computed in one-loop
perturbation theory in ref.~\cite{mms}. Notice that we have preferred
to express $\bar h\vec D^2_s h $ in terms of the subtracted 
Lagrangian operator,
which explicitly contains the residual mass $\delta m$. In this way we
can use the equations of motion of the subtracted Lagrangian.
This will prove useful below.
\par
 In order to eliminate the quadratic and linear  power divergences,
a possible  non-perturbative  renormalization condition for 
 $\bar h\vec D^2_s h $  is that its subtracted matrix element, computed
 for a quark at rest in the Landau gauge, vanishes:
\be 
\langle h(\vec p=0) \vert \bar h\vec D^2_s h  \vert h(\vec p=0) \rangle=0
\, . \label{vanish}
\ee
This is equivalent to defining the subtraction constants through the
relation (in the following we will work in lattice units, setting
$a=1$)
\beq
\rho_{\vec{D}^{2}}(t)  = c_1\,+\,c_2\, t ,\label{eq:twenty}\eeq
where
\beqn \rho_{\vec{D}^{2}}(t) \equiv   
\frac{\sum_{t^\prime=0}^{t}\sum_{\vec{x},\vec{y}}\,
 \langle S(\vec x,t\vert \vec y,t^\prime)
 \, \vec D^2_y(t^\prime)\, S(\vec y,t^\prime\vert \vec 0,0)\rangle}{ 
\sum_{\vec{x}}\,\langle S(\vec x,t\vert \vec 0,0)\rangle}\, , \label{eq:c12}
\eeqn
where $y=(\vec y, t^\prime)$.
$S(\vec x,t\vert  \vec 0,0)$ is the heavy-quark propagator  between
the point of coordinates $(\vec x,t)$ and $(\vec x=0,t=0)$ 
computed on a single
gauge field configuration;
 $\langle \dots \rangle$ denotes the average over the gauge field 
configurations. 
\par 
For most applications it is only the constant $c_2$ that is
required; $c_2$ can also be determined  directly by eliminating the sum over
$t^\prime$ in eq.~(\ref{eq:c12}):
\be
c_2 =\rho_{\vec{D}^{2}}(t^\prime,t) =\frac{\sum_{\vec x,\vec y}\,
 \langle S(\vec x,t\vert \vec y,t^\prime)
 \, \vec D^2_y(t^\prime)\, S(\vec y,t^\prime\vert \vec 0,0)\rangle}{
\sum_{\vec{x}}\,\langle S(\vec x,t\vert \vec 0,0)\rangle}
\label{eq:c2tp}\ee
for $t^\prime\neq 0, t$. The term proportional to $c_1$ disappears because,
by the equations of motion of the subtracted Lagrangian, 
it can only contribute to contact terms.
\par  
The renormalized kinetic energy $\lambda_1$ of the $b$-quark inside a $B$-meson
is then given by
  \be \lambda_1  =\, Z_{\vec D^2_s}
\left(\lambda_1^{{\rm bare}} 
- c_{2}\right) =\,  \left( 1- \frac{\alpha_s}{4\pi}X_{\vec D^2_s}\right)
\left(\lambda_1^{{\rm bare}} 
- c_{2}\right) \, ,
\label{eq:nextor}
\ee
where  $\lambda_1^{{\rm bare}}=\langle B
\vert \bar h \vec{D}^{2} h \vert B \rangle/(2 M_B)$.
The term proportional to $X_{\vec D^2_s}$ in eq.~(\ref{eq:nextor}) is
absent in continuum formulations of the HQET, and is a manifestation
of the lack of reparametrization invariance in its lattice version. It
has been calculated in ref.~\cite{mms}. One can argue that the
counter-term $c_2$ defined above is gauge invariant, in spite of the
fact that it has been derived from quark Green functions in a fixed
gauge.  As a consequence, since the matching constant $ Z_{\vec
  D^2_s}$ is gauge invariant, $\lambda_1$ defined in
eq.~(\ref{eq:nextor}) does not depend on the gauge.

\par $\lambda_1^{{\rm bare}}$ can
be determined from a computation of two- and three-point $B$-meson
correlation functions in the standard way. Given the quantities
\beqn
C_{B}(t) &=& \sum_{\vec x}
\langle 0|J_B(\vec x, t)\  J^\dagger_B(\vec 0,0)|0\rangle
\quad \mbox{and} \nonumber \\
C_{\vec D^2}(t^\prime, t) &=& \sum_{\vec x, \vec y}
\langle 0|J_B(\vec x, t)\ \bar h(\vec y,t^\prime)\vec D_y^2(t^\prime) h(\vec 
y,t^\prime) 
J^\dagger_B(\vec 0,0)|0\rangle
\, ,\label{eq:cd2s}\eeqn
we have, for sufficiently large values of $t^\prime$ and $t-t^\prime$: 
\beqn &\,&
C_{B}(t)\ \rightarrow Z^2 \, \exp
\left(-({\cal E}-\delta \overline{m})t\right)\, , \nonumber \\ &\,&
C_{\vec D^2}(t^\prime, t)\ \rightarrow Z^2 \, \lambda_1^{{\rm bare}}\,\exp
\left(-({\cal E}-\delta \overline{m})t\right)\, ,
\label{eq:cd2sasymp}\eeqn
where $J^\dagger_B(\vec x, t)$ ($J_B(\vec x, t)$) is a
source which creates (annihilates) a pseudoscalar $B$-meson state
from the vacuum; ${\cal E}$ is the ``bare" binding energy and 
$\delta \overline{m}=\delta \overline{m}(\delta m)$ a suitable
mass counter-term \cite{cgms,ms,gms}.
A convenient way to extract $\lambda_1^{{\rm bare}}$ is to compute
the ratio
\beq
R_{\vec D^2}(t^\prime, t)=\frac{C_{\vec D^2}(t^\prime, t)}{C_B(t)}\rightarrow
\lambda_1^{{\rm bare}}\ .
\label{eq:deltaasymp}\eeq
As usual $\lambda_1^{{\rm bare}}$ must be evaluated in an interval in which 
$R_{\vec D^2}(t^\prime, t)$ is independent of the times 
$t^\prime$ and $t$, so that the
contribution from  excited states and contact terms can be neglected.
\par The  computation of the 
chromo-magnetic matrix element $\lambda_2$, defined as
\be \label{eq:lambda2}
\lambda^{{\rm bare}}_2 = 
\frac{1}{3}\langle B \vert \bar h\, \frac{1}{2}\, \sigma_{i
j}\, G^{i j}\, h \vert B \rangle/(2\, M_{B}) \, ,
\ee 
proceeds in a similar   way. The most important difference  
is  that $\lambda_2$ is free of power divergences
so that, at least in principle,
there is no need of a non-perturbative subtraction
\be \lambda_2 = Z_{\vec{\sigma}\cdot \vec{G}} \,   \lambda^{{\rm bare}}_2\, ,
 \label{eq:l2r} \ee
where $Z_{\vec{\sigma}\cdot \vec{G}}$ is the renormalization constant necessary
to remove the logarithmic divergence present in the bare operator \cite{fh}.
In eq.~(\ref{eq:lambda2}), $\sigma_{i j}=1/2\, [\gamma_{i},\gamma_{j}]$
is the spin Dirac matrix and $G^{i j}$ is the colour field-strength tensor.
In our computation we use the clover  definition of
$G^{i j}$ introduced in ref.~\cite{mandula}. 
To determine $\lambda_2$, we also used a  ratio of
correlation functions as in eq.~(\ref{eq:deltaasymp}), with
the chromo-magnetic operator $1/2\ \vec{\sigma}\cdot \vec{G}$
 inserted in   the three-point function.  In this case  we will denote the
ratio (and the corresponding three-point
function) as $R_{\vec{\sigma}\cdot \vec{G}}(t^\prime, t)$
($C_{\vec{\sigma}\cdot \vec{G}}(t^\prime, t)$).

\section{Numerical calculation of $\lambda_1$} 
\label{numerical1}
As explained in the previous section, the renormalization procedure,
which defines a finite kinetic energy operator, requires the
computation of $c_1$ and $c_2$.  These subtraction constants were
obtained using a sample of $600$ configurations on a $24^{3}\times 40$
lattice at $\beta=6.0$.  Also $\lambda_1^{\rm bare}$ has been computed on
the same set of gauge field configurations as $c_1$ and $c_2$. The
relevant two- and three-point meson correlation functions were
computed using the improved SW-Clover action \cite{clover} for the
light quarks (with improved-improved propagators \cite{tassos}), in
the quenched approximation, at three values of the light-quark masses
corresponding to $K=0.1425$, $0.1432$ and $0.1440$, where $K$ is the
Wilson parameter. As in our previous studies of the lattice HQET
\cite{alltonw}, for the heavy meson sources we used the standard axial
currents smeared over cubes of size $L_S=5$, $7$ and $9$.  All the
errors have been computed with the jack-knife method by decimating 12
configurations at a time.
\par The same  600 configurations, quark propagators and smeared 
sources were used for the calculation
of both $\lambda_1$ and   $\lambda_2$. Thus we will 
not repeat the discussion of the parameters of the numerical calculation
when discussing the results for $\lambda_2$.
\subsection{Determination of the subtraction constant $c_{2}$} 
\label{sec:c12} 
\begin{figure} \vspace{9pt}
\begin{center}\setlength{\unitlength}{1mm} \begin{picture}(160,100)
\put(0,-58){\includegraphics{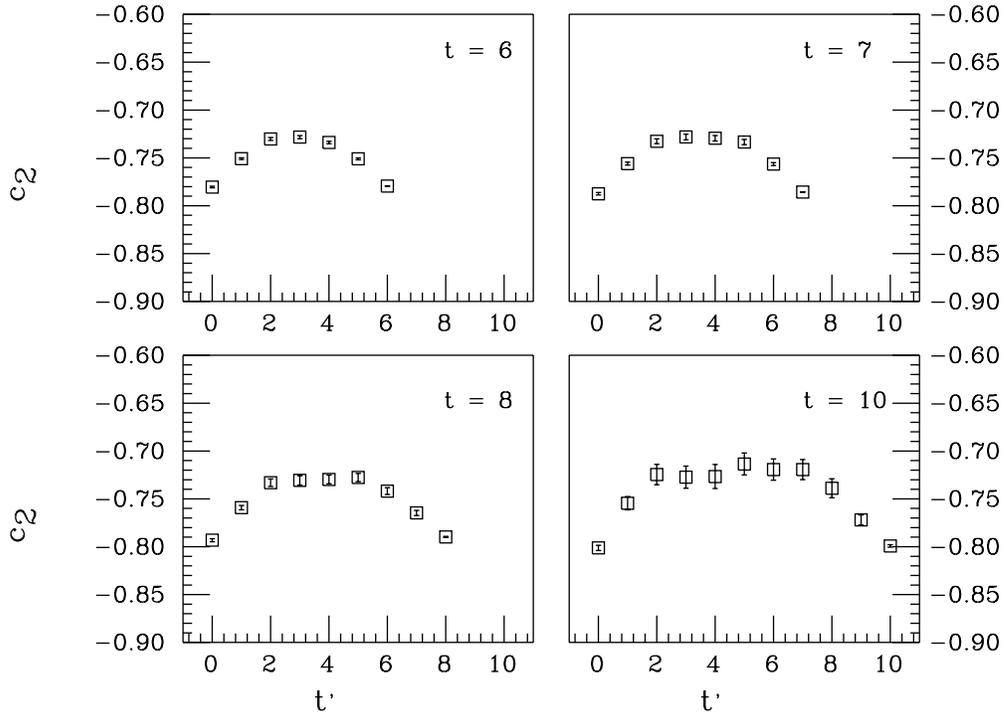}} \end{picture} \end{center}
\caption{\it{The ratio $\rho_{\vec{D}^{2}}(t^\prime,t)$ defined in 
eq.~(\protect\ref{eq:c2tp}), at $\beta=6.0$,
as a function of $t^\prime$,  at several values of $t$, $t=6,7,8$ and $10$.}}
\label{fig:c2tp}
\end{figure}

In order to compute the renormalized  $\lambda_1$, we only need to determine
the subtraction constant $c_2$, cf. eq.~(\ref{eq:nextor}). There are several
methods to define and extract $c_2$ from the heavy-quark Green functions:
\begin{enumerate} 
\item   One possibility is to fit the time-dependence 
of $\rho_{\vec{D}^{2}}(t)$  at large time distances  ($t \to \infty$) to
a straight line, see eq.~(\ref{eq:twenty}).  This corresponds to imposing the
renormalization condition (\ref{vanish}) at  $p_4=0$. 
Previous studies of the 
propagator at large time distances \cite{cgms,gms}, and the results
presented below,  suggest that 
 the infrared limit of the ratio in eq.~(\ref{eq:c12})
does indeed exist, in spite of  the confinement effects
associated with the heavy-quark propagator. 

\item Similarly, it is possible to define $c_2$ from
$\rho_{\vec{D}^{2}}(t^\prime , t)$, see eq.~(\ref{eq:c2tp}),
 in the limit $t^\prime \to \infty$ and $t-t^\prime  \to \infty$.
\item A different possibility consists in defining $c_2$ at fixed
  $t^\prime \neq 0$ and $t-t^\prime \neq 0 $ (for sim\-pli\-ci\-ty we
  only discuss the case $t^*=t^\prime=t-t^\prime$). We denote the
  corresponding subtraction constant by $c_2(t^*)$:
  $t^*$ parametrizes the renormalization prescription
  dependence and can be considered as the renormalization point in
  coordinate space.  Since the lattice kinetic energy operator has no
  anomalous dimension, we do not expect any
  dependence of $c_2(t^*)$ on the renormalization point $t^*$ at any
  order of perturbation theory, and this is supported by our numerical
  results \footnote{ This is true when $t^*$ is chosen in such a way
    as to avoid contact terms.}.  The use of two- and three-point
  Green fuctions at small times, $t^*\Lambda_{QCD}\ll 1$, to define
  $c_2$ and hence $\lambda_1$ does not require any assumption about
  the behaviour of the heavy-quark propagator at large times.  Our
  numerical results do not show any visible dependence (up to contact
  terms) on $t^*$.
\item On a lattice with a finite lattice spacing, contact terms are
  not localized on a single point, $t^\prime=0$ or $t^\prime=t$, but
  are smeared over several lattice spacings (see for example
  ref.~\cite{ward} and fig.~\ref{fig:c2tp}, to be discussed below).
  For this reason, besides $\rho_{\vec{D}^{2}}(t)=
  \sum_{t^\prime=0}^{t} \rho_{\vec{D}^{2}}(t^\prime , t)$, in order to
  reduce the effect of contact terms in the extraction of $c_2$, we
  have also linearly fitted the quantity \be
  \rho^{\Delta}_{\vec{D}^{2}}(t)= \sum_{t^\prime=\Delta}^{t-\Delta}
  \rho_{\vec{D}^{2}}(t^\prime , t) = c_1 +c_2 t \, , \ee where the
  points close to $t$ and to the origin have been eliminated from the
  sum.
\end{enumerate}
\vskip 0.3 cm
In fig. \ref{fig:c2tp}, we plot $\rho_{\vec{D}^{2}}(t^\prime, t)$,
 as defined in
eq.~(\ref{eq:c2tp}), as a function of  $t^\prime$, at fixed $t=6,7,8$
and $10$.
Up to contact terms, we expect $\rho_{\vec{D}^{2}}(t^\prime,t)$ to be a
constant in $t^\prime$, at fixed $t$, and also to be 
independent of
$t$. If the contact terms were entirely due to the mixing of the
kinetic energy operator with the inverse propagator, as in
eq.~(\ref{eq:d2ren}), we should find two spikes, at $t^\prime=0$  and
$t^\prime=t$, and a constant value of $\rho_{\vec{D}^{2}}(t^\prime,t)$ 
for $t^\prime \neq 0,t$. The presence of operators of higher dimension,
due to discretization errors, introduces terms that behave as
derivatives of $\delta$-functions (in time), giving rise to the 
bell-shape behaviour of $\rho_{\vec{D}^{2}}(t^\prime,t)$ shown in 
fig.~\ref{fig:c2tp}. Thus in order to obtain $c_2$, we have to look for a
plateau in the  central region in $t^\prime$, at large values of $t$.
>From the figure, we see that, since the contact terms
extend over about two lattice spacings in both $t^\prime$ and $
t-t^\prime$, it is possible to recognize a plateau in
$t^\prime$ only for $t=7$--$10$ (at larger values of $t$ the statistical
errors are too large to draw a firm conclusion).  At $t=6$ the best that we can
do is to consider only the point at $t^\prime=3$.
To obtain $c_2$ we have made a weighted average of the values of
$\rho_{\vec{D}^{2}}(t^\prime,t)$ over the plateau region, at fixed $t$.
The results are  given in table \ref{tab:c2tp}. 
\begin{table} \centering
\begin{tabular}{||c|c|c|c||}
\hline
\hline
\multicolumn{4}{||c||}{$c_2$ from $\rho_{\vec{D}^{2}}(t^\prime,t)$
and $\rho^{\Delta=2}_{\vec{D}^{2}}(t)$}\\
\hline \hline
$t$&
\multicolumn{1}{c|}{{\it Interval}}&
\multicolumn{1}{c|}{$\chi^2$}&
\multicolumn{1}{c||}{$c_2$}\\
\hline \hline
6& 3--3&0.0&$-$0.728(2) \\
7& 3--4&0.1&$-$0.729(3) \\
8& 3--5&0.1&$-$0.729(4) \\
9& 3--6&0.4&$-$0.727(6) \\
10& 3--7&0.2&$-$0.721(10) \\
\hline \hline
\multicolumn{3}{||c|}{{\it Interval}}&
\multicolumn{1}{c||}{$c_2$}\\
\hline \hline
\multicolumn{3}{||c|}{$(6:7)$--$(10:16)$}&
$-$0.727(6) \\
\hline\hline
\end{tabular}
\caption{\it{Values of $c_2$ obtained as explained in the text.
We also give  the uncorrelated $\chi^2$.}}
\label{tab:c2tp}
\end{table}
In the table, the column 
 {\it Interval} denotes the interval in $t^\prime$ over
which $\rho_{\vec{D}^{2}}(t^\prime,t)$ has been averaged.

\par If we fit $\rho_{\vec{D}^{2}}(t)$  to a straight line 
on intervals that include small values of $t$, the presence of
extended, time-dependent contact terms can also induce a systematic
error in the determination of $c_2$ from this quantity.  For this
reason, in the extraction of $c_2$ from a fit to
$\rho_{\vec{D}^{2}}(t)$ we only used
$\rho^{\Delta=2}_{\vec{D}^{2}}(t)$ with $t \ge 6$. The result is also
given in table \ref{tab:c2tp}. For this quantity {\it Interval}
denotes the time interval on which $\rho^{\Delta=2}_{\vec{D}^{2}}(t)$
has been fitted and $(6:7)$--$(10:16)$ indicates minimum--maximum
values of the time on which the fit has been made, i.e. the fits were
obtained starting with $t=$ 6 or 7 and ending at any of the points
between 10 and 16.

\par  The results given in the table show that the values 
of $c_2$ as obtained from
$\rho_{\vec{D}^{2}}(t^\prime,t)$ are essentially independent of $t$
(and of $t^\prime$), as expected from the vanishing of the anomalous 
dimension, and are indistinguishable from the value 
extracted with  a linear fit to   $\rho^{\Delta=2}_{\vec{D}^{2}}(t)$. 
Our best estimate for this quantity is then
\be c_2= -0.729 \pm 0.005 \, , \label{eq:c2res} \ee
to be compared with our previous result $c_2=-0.73 \pm 0.02$
\cite{cgms}. In eq.~(\ref{eq:c2res}), the central value has been taken
from $\rho_{\vec{D}^{2}}(t^\prime=3$--$5, t=8)$; the error
includes the  fluctuations 
of $\rho_{\vec{D}^{2}}(t^\prime,t)$ between different points
in $t^\prime$, at fixed $t$, as well as the variation of 
$\rho_{\vec{D}^{2}}(t^\prime,t)$ with $t$ or variations 
of $c_2$ as derived from 
a fit to $\rho^{\Delta=2}_{\vec{D}^{2}}(t)$ with respect 
to  different time intervals.

\subsection{Determination of  $\lambda^{{\rm bare}}_1$} 
\label{sub:lambda1} 
 
The procedure to extract operator matrix elements is standard. It is
the same as the second method that we have  used in the previous subsection
to determine $c_2$. At fixed $t$, we study the behaviour of the ratio
$R_{\vec D^2}(t^\prime, t)$ as a function of $t^\prime$,  searching for a  plateau
in $t^\prime$. Here $\lambda_1^{{\rm bare}}$ is defined by  the  weighted 
average of the data points in the central  plateau region, if this
exists. We will take as our best determination of  $\lambda_1^{{\rm
bare}}$,  the value  evaluated in a time interval where the ratio 
$R_{\vec D^2}(t^\prime, t)$ appears to be independent of both $t$ and $t^\prime$.
In addition, we have to require  that the  lightest state has been
isolated. With the smeared sources used in the present case, we know
that this happens at time distances $(t-t^\prime)$ and $t^\prime 
\ge 4$--$5$.  This implies that the total time distance
$t$ for $R_{\vec D^2}(t^\prime, t)$ has to be at least $8$--$10$. Moreover,
by using $(t-t^\prime)$ and $t^\prime  \ge 4$--$5$, we eliminate the
contact terms which, on the basis of the discussion in the previous
subsection,  are expected to be present up to
distances of order $2$--$3$.
\par In the two- and three-point meson correlation functions,
in order to improve the isolation of the lightest meson state at short
time distances, we have used, as in our previous studies \cite{alltonw},
 the following sources
\begin{eqnarray}
J^{S}_{B}(x) &=& \sum_{i} \bar{h}(x_{i})\,\gamma_{0}\, \gamma_{5}\, q(x)\\
J^{D}_{B}(x) &=& \sum_{i,j} \bar{h}(x_i)\,\gamma_{0}\, \gamma_{5}\, q(x_j)
\, ,\end{eqnarray}
where we sum the position of the fields $x_i$
over cubes of sizes $L_S=5,7$ and $9$, centred on $x$. 
\par
\begin{figure}
\vspace{9pt}
\begin{center}\setlength{\unitlength}{1mm}
\begin{picture}(160,60)
\put(0,-55){\includegraphics{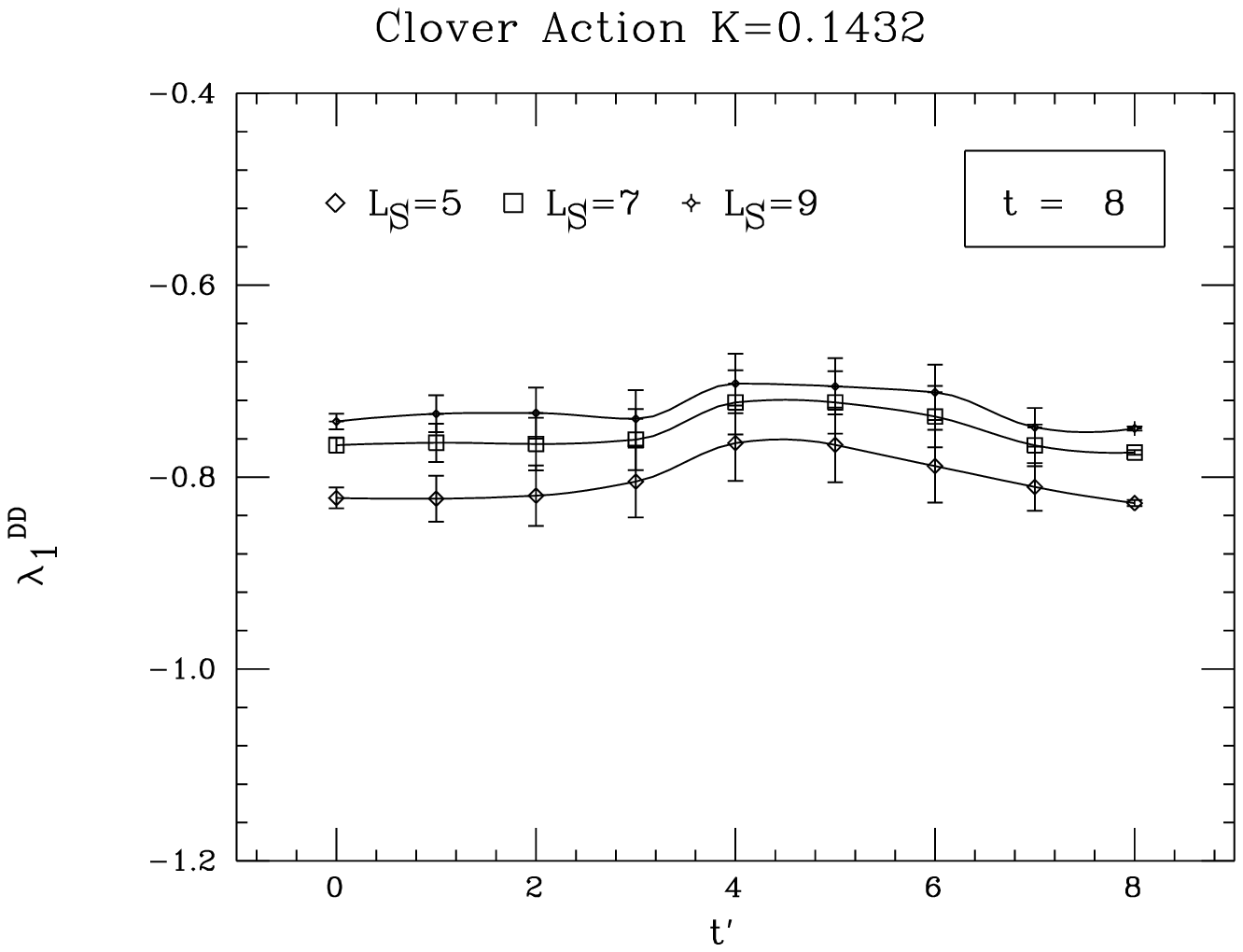}}
\end{picture}
\end{center}
\caption{\it{The ratio $\lambda_1^{DD}=R_{\vec{D}^{2}}(t^\prime, t)$ 
as a function of $t^\prime$, at $t=8$, for smearing sizes $L_S=5,7$ and $9$.}}
\label{fig:smea}
\end{figure}

 In order to show the dependence of the results
on $L_S$, we give in fig.~\ref{fig:smea} the results 
for $\lambda_1^{DD}=R_{\vec D^2}(t^\prime, t)$ as a function 
of $t^\prime$, at fixed $t=8$,
obtained using the double-smeared sources $J^{D}_{B}(x)$, at the  value
of the light-quark Wilson parameter $K=0.1432$.
We observe some  dependence of $\lambda_1^{DD}$ on $L_S$. For $L_S \le 5$,
this is not surprising, since we know, from previous studies
of smeared-source correlators, that it is not possible to have a good
isolation of the lightest state for $L_S \le 5$. Between $L_S=7$ and
$L_S=9$, there is a small residual shift in the
value of $\lambda_1^{DD}$ which will be taken into account in the evaluation
of the error on $\lambda_1^{{\rm bare}}$. In the following, we will use
$L_S=7$ with double-smeared sources
(which was found to be the optimal cubic smearing~\cite{alltonw}) to determine 
the central value of $\lambda_1^{{\rm bare}}$.
\par
\begin{figure} \vspace{9pt}
\begin{center}\setlength{\unitlength}{1mm}
\begin{picture}(160,60)
\put(0,-55){\includegraphics{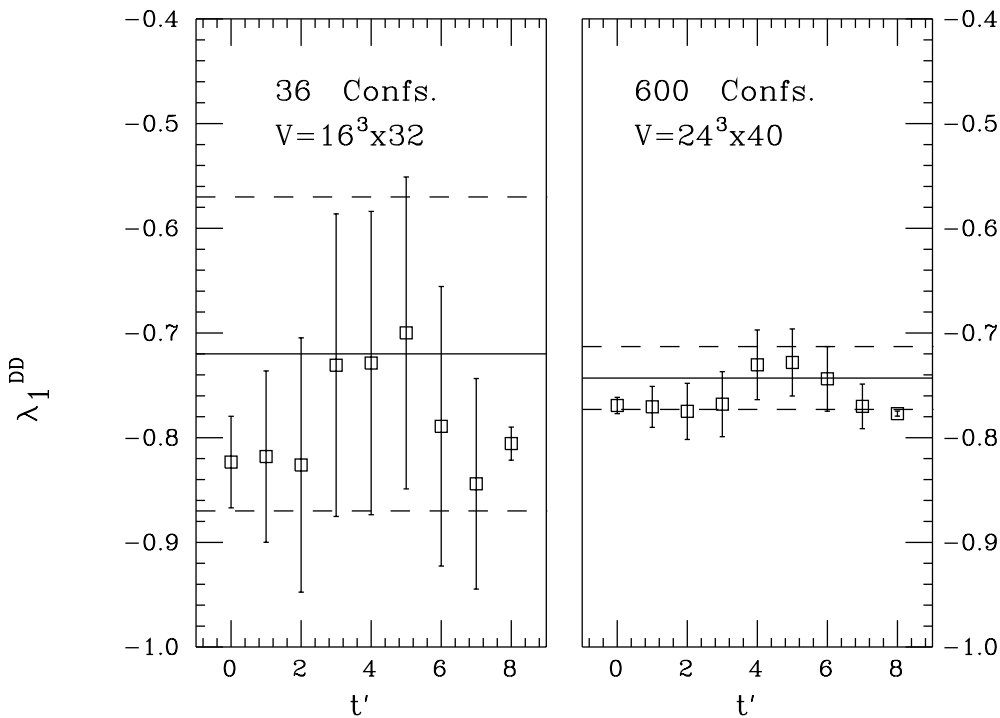}}
\end{picture}
\end{center}
\caption{\it{The ratio
$\lambda_1^{DD}=R_{\vec{D}^{2}}(t^\prime, t)$  as a function of
$t^\prime$, at $t=8$ for $L_S=7$. The results obtained in ref.~[1]
and those  of the present study
are shown on the left-hand  and right-hand side 
of the figure respectively.}}
\label{fig:confro} \end{figure}

 In fig.~\ref{fig:confro}, we give, for $L_S=7$ and 
at $K=0.1425$, the
plateau for $\lambda^{DD}_1$ as a function of $t^\prime$ at $t=8$. In
the figure, the extracted value of $\lambda^{{\rm bare}}_1$, obtained
by a weighted average of $\lambda^{DD}_1$ in the interval
$t^\prime=3$--$5$ (the same as the corresponding interval used to
determine $c_2$) is also given (solid line). The band limited by the
dashed lines corresponds to the error of this quantity. For comparison
we also report the results obtained in ref.~\cite{cgms}, at the same
value of $\beta$, on a volume $16^3 \times 32$, with a statistics of
only $36$ configurations.
The significant reduction of the statistical errors with respect to
our previous study allows for a much better identification of the
plateau region, which had been assumed to exist, but could not be
really observed, in ref.~\cite{cgms}.  Morevover, in the present case,
we have computed the meson correlation functions at three values of
the hopping parameter $K$, which allows us to extrapolate in the mass
of the light quarks $\lambda_1^{{\rm bare}}$ to the chiral limit
($K_c=0.145431(10)$), corresponding to the kinetic energy of the heavy
quark in the $B_d$ meson.
\par Another method to extract $\lambda_1^{{\rm bare}}$ is to
introduce the quantity
\be R_{\vec{D}^{2}}(t)= \sum_{t^\prime=0}^{t} 
R_{\vec{D}^{2}}(t^\prime, t) = \alpha_1 + \lambda_1^{{\rm bare}} t 
\, , \label{sumr} \ee
and to fit it to a straight line. This method was 
used in old  numerical calculations of hadronic matrix elements,
such as  the kaon  $B$-parameter \cite{mm} and  the nuclueon $\sigma$-term
\cite{marie}, and more recently by the UKQCD collaboration for the 
calculation of $\lambda_2$ \cite{ukqcdl2}, see also below. 
A possible inconvenience of this method 
is due to the fact that  the presence of contact terms,
and excited states, when $t$ is not large enough
 can significantly distort the value
of the slope in time, and hence of the value of the matrix element 
extracted from the
correlation function. 
In this respect, the use of $R_{\vec{D}^{2}}(t^\prime, t)$ 
is better in that it allows us  to verify the existence of a plateau, and
the independence of the results of $t$ and $t^\prime$. 
\begin{table} \centering
\begin{tabular}{||c|c|c||}
\hline
\hline
\multicolumn{3}{||c||}{$\lambda_1^{{\rm bare}}$ from 
$R_{\vec{D}^{2}}(t^\prime,t)$
and $R_{\vec{D}^{2}}(t)$ for $L_S=7$}\\
\hline \hline
$K$&
\multicolumn{1}{c|}{$t=8$, $t^\prime=3$--$5$}&
\multicolumn{1}{c||}{Fit with  $6 \le t \le 16$} \\
\hline \hline
0.1425& $-0.74(3)$&$-0.70(4)$ \\
0.1432& $-0.74(3)$&$-0.70(4)$ \\
0.1440& $-0.73(3)$&$-0.69(4)$ \\ \hline
0.145431& $-0.71(3)$&$-0.67(4)$ \\
\hline\hline
\end{tabular}
\caption{\it{Values of $\lambda_1^{{\rm bare}}$ obtained as explained in the text.}}
\label{tab:l1bareb}
\end{table}
\par We  give in table \ref{tab:l1bareb} the results obtained by using the two
methods outlined above.  From this table, we see that the typical statistical
error is of the order of $0.03$. Taking into account the fluctuations of the
results between different points
in $t^\prime$, at fixed $t$, and in $t$, of the variations 
between results obtained with the linear fit to
$R_{\vec{D}^{2}}(t)$ on  different time intervals,
and of the differences between $L_S=7$ and $L_S=9$, we estimate,
for the value extrapolated to the chiral limit: 
\be \lambda_1^{{\rm bare}}=-0.69 \pm 0.03 \pm 0.03 \, .\ee
In order to appreciate the reduction of the errors, we note that our
old result was $\lambda_1^{{\rm bare}}=-0.72 \pm 0.14$ at 
$K=0.1425$~\cite{cgms}.
\subsection{Determination of the ``physical" value of 
$\lambda_1$} \label{kinener}
We are now in a position to  compute the ``physical" value of
$\lambda_{1}$, i.e. the finite matrix element of the subtracted operator. In
this section   the lattice  spacing $a$ is written explicitly in the formulae. 
For the physical value of the inverse lattice spacing  we  use 
 $a^{-1} = 2.0 \pm\, 0.2$ GeV, where the uncertainty takes into account the
spread of values obtained in the quenched approximation from different
determinations of $a$ (from the string tension, from the mass spectrum, 
from $f_\pi$, etc.).
For the subtraction  constant $c_2$,  we use the result given in 
eq.~(\ref{eq:c2res})
 \be a^2 \, c_2\, =\, -  0.729\, \pm \, 0.005  \, ,\ee
together with 
\be a^2 \, \lambda_1^{{\rm bare}}=-0.69 \pm 0.03 \pm 0.03 \, .\ee
We thus obtain
\be  \lambda_{1}\, =\, a^{-2}\, Z_{\vec D_S^2}\, \Bigl( 
a^2 \lambda_{1}^{{\rm bare}}\,
-  \, a^2 c_2\Bigr)  =\, 0.18\, \pm \,0.22 \quad \mbox{GeV}^2\ .
\ee
We have taken   $Z_{\vec D_S^2} = 1.13$  obtained from the expression of 
the one-loop renormalization constant \cite{mms}
\be Z_{\vec D_S^2} \, =\, 1\, +\, \left(\frac{\alpha_{s}}{\pi}\right)\,
\frac{C_{F}}{6}\, \frac{1}{2 \pi}\, \int_{-\pi/a}^{\pi/a}\,
d^{3} k\, \frac{1}{\sqrt{(1+A)^{2}-1}} \, ,\label{zz}
\ee where \be A\,=\, \sum_{i=1}^{3}\, (1-\cos(k_{i} a))\, ,\ee
and used  the boosted strong coupling constant \cite{lpmk}
\be
\alpha^{latt}_{s}\, =\, \frac{6}{4 \pi \beta u_{0}^{4}} \, ;
\ee
$u_{0}$ is a measure of the average link variable for which we take
$u_{0}=(8 K_{c})^{-1}$ with $K_{c}=0.145431(10)$.  The value of the
renormalization constant in standard perturbation theory would be
$Z_{\vec D_S^2} = 1.07$. One can try to reduce the error by taking the
difference between $\lambda_{1}^{{\rm bare}}$ and the counter-term $c_2$,
for each jack-knife individually. In this case we obtain
\be
\lambda_1 = \, 0.09\, \pm \,0.14 \quad \mbox{GeV}^2\ ,
\label{eq:l1best}\ee
which we take to be our best result.

\par One could argue that the subtraction is not really necessary, since the
effective theory on the lattice does not have renormalons.  Even
though this is indeed true, the difficulty in the determination of
corrections of order $1/m_Q$ related to the kinetic energy operator
would remain the same. The argument goes as follows. The bare kinetic
energy operator has a very large matrix element $a^{-2} \times (a^2
\lambda_1^{{\rm bare}}) \sim 2^2 \times -0.69$~GeV$^2=-2.8$ GeV$^2$,
while one expects a correction due to the kinetic energy of the heavy
quark of the order of the squared Fermi momentum $p_F^2 \sim
\Lambda_{{\rm QCD}}^2 \sim 0.1$--$0.6$~GeV$^2$. Thus the huge
contribution of the matrix element of the bare operator has to be
compensated by the corresponding term in the coefficient function of
$\bar h h$~\cite{ms2}.  This requires an extreme accuracy in the
perturbative calculation of the coefficient function and it remains
true in the subtracted as well as in the unsubtracted case \footnote{
  Similar problems are encountered also in different approaches to
  determine $\lambda_1$, see for example ref.~\cite{extract2}.}.
\par In the past $\lambda_1$ has been computed in different models 
and using QCD sum rules~\cite{neubertl1}--\cite{qcdsr}. Bounds on
$\lambda_1$ were also obtained from zero-recoil sum
rules~\cite{bounds}, which however are weakened by higher-order
perturbative corrections \cite{bounds2}.  The extraction of its value
from experimental data has also been attempted
\cite{zn}--\cite{extract11}; small values of $\lambda_1$ are suggested
by the theoretical analysis of ref.~\cite{virial} and by the recent 
studies, \cite{extract2, chernyak}.  
The results of the different theoretical estimates
are given in table \ref{tab:esti}~\footnote{The value attributed to 
ref.~\protect\cite{es} in table~\protect\ref{tab:esti}, was  
extracted from the results of this paper by Neubert in 
ref.~\cite{neubertprep}. The value for ref.~\protect\cite{extract2}
was taken from the revised version of this paper.}.
\begin{table} \centering
\begin{tabular}{||c|c|c||}
\hline \hline
Reference &
\multicolumn{1}{c|}{Method}&
\multicolumn{1}{c|}{$-\lambda_1$ (GeV$^2$) }\\
\hline \hline
Eletsky and Shuryak \protect\cite{es}& QCD sum rules &$0.18\pm 0.06$  \\
Ball and Braun \protect\cite{qcdsr}& QCD sum rules &$0.52\pm 0.12$  \\
Bigi et al. \protect\cite{bounds}& ZR sum rules &$\ge 0.36$  \\
Kapustin et al. \protect\cite{bounds2}& 
ZR sum rules +$O(\alpha_s)+O(\alpha_s^2 \beta_0)$
& No Bound  \\
Ligeti and Nir \protect\cite{zn}& 
Experiment&$\le 0.63$ if $\labar \ge 240$ MeV  \\
 & Experiment&$\le 0.10$ if $\labar \ge 500$ MeV  \\
Gremm et al. \protect\cite{extract2}& Experiment&$0.19\pm 0.10$  \\
Chernyak \protect\cite{chernyak} & Experiment & $0.14 \pm 0.03$ \\ 
\hline\hline
\end{tabular}
\caption{\it{Some of the 
values of  $\lambda_1$ obtained in different theoretical
analyses. ``Experiment" denotes the extraction of $\lambda_1$ from the
experimental data, for example the charged-lepton spectrum distribution
in semileptonic $B$-meson decays. ``ZR sum rules" denotes the zero recoil
sum rules.}}\label{tab:esti}
\end{table}
\par A  comparison of the different results and limits given in table
\ref{tab:esti} between them and with ours is difficult for two reasons
that we briefly discuss. We have seen that the subtraction of the
quadratic divergence is necessary in order to define a kinetic energy
operator, the matrix elements of which are finite.  The precise
definition of the renormalized operator, and hence the value of
$\lambda_1$, depends on the renormalization procedure and different
definitions have been used in the results of table \ref{tab:esti}. For
example in ref.~\cite{extract2} the definition of $\lambda_1$ is
implicitly given by the weight function $W_\Delta$, which depends
quadratically on the ``ultraviolet cut-off" $\Delta$, similarly to the
quadratic dependence of the lattice $\lambda^{{\rm bare}}_1$ on the
inverse lattice spacing.  The relation between the different
definitions is in principle computable in perturbation theory,
although this exercise has not been done to date. The second, related
remark (which is general for power-suppressed corrections in all
effective theories) concerns the precision that can be achieved in
evaluating, for a given physical quantity, the power corrections at a
given order in the inverse quark mass $1/m_Q$.  As discussed in
ref.~\cite{ms2}, there are essentially two procedures to compute the
power corrections. One is based on the determination of the parameters
of the HQET, such as $\labar$ and $\lambda_1$, from the measurement of
some experimental quantity, as was done for example in
ref.~\cite{extract2}.  The parameters found with this method can then
be used to predict other physical quantities. The other approach is
that followed in this paper.  The parameters of the HQET are defined
and computed with some non-perturbative technique, the lattice
numerical simulations in our case, and then used in the calculation of
physical quantities, with the same renormalization scheme. In both
cases, however, it turns out to be rather difficult to achieve a
sufficient precision in the calculation of the power corrections,
because of the truncation of the perturbative series for the Wilson
coefficients of the $1/m_Q$ expansion.  The truncation of the
perturbation series limits also the possibility of accurate
comparisons between values of $\labar$ and $\lambda_1$ obtained in the
different approaches. A detailed discussion of this problem, with
several examples, can be found in ref.~\cite{ms2} (see also
\cite{ms}).  The disappearance of the bound found in ref.~\cite{bigi}
due to higher-order perturbative corrections, which are quadratically
divergent in the cut-off ($\sim \alpha_s \Delta^2$) \cite{bounds2}, is
just an explicit example of what is expected on the basis of the
general arguments of refs.~\cite{ms,ms2}.  This holds true also in the
case of ref.~\cite{extract2}, where the one-loop perturbative
corrections to the quantities used to obtain $\labar$ and $\lambda_1$
are very small. For example, when using the results of this analysis
to predict the mass of the $b$-quark in the $\overline{MS}$ scheme, an
intrinsic error of about $200$ MeV from higher-order perturbative
contributions is expected \cite{ms2}, much larger than the correction
given by the kinetic energy operator, which is of the order of 30 MeV.
For the above reasons, although some of the results in table
\ref{tab:esti} look barely compatible with each other and with our
result, the only sensible thing to do is to compare the predictions of
the different approaches for some physical quantity (some width, the
$\overline{MS}$ mass, etc.)  to the desired accuracy in $1/m_Q$ (or
$1/m_Q^2$), and subject to the condition that the perturbative
corrections are under control.  The present differences can be taken
as a demonstration that these higher-order corrections are likely to
be considerable.

\par Although $\lambda_1^{{\rm bare}}$ for a single hadron is not a physical
parameter, the difference of the $b$-quark kinetic energy of two
different hadrons is a well-defined quantity which does not require
any subtraction. For example, we can study the kinetic energy in a
$B$-meson as a function of the mass of the light quark in the meson,
by fitting
\beq \lambda_1^{{\rm bare}}= A_{\lambda_1} + B_{\lambda_1} M_{PS}^2 \,
, \eeq
where $ M_{PS}^2$ is the mass of the pseudoscalar meson made by
two  light quarks with the same mass as the light quark in the 
$B$-meson. From our data, by assuming that $ M_{PS}^2$ depends linearly
on the value of the quark masses, we get
\beq \lambda_1(B_s)-\lambda_1(B_d)= - 0.09 \pm 0.04 \,\, \mbox{GeV}^2 ,
\label{eq:09}\eeq
where we have also included the factor due to the renormalization 
constant of the lattice bare operator. The result in eq.~(\ref{eq:09})
compares well with the number that can be extracted from the following
combination of heavy-meson masses:
\beq \lambda_1(B_s)-\lambda_1(B_d)=
\frac{\overline{M}_{B_s} - \overline{M}_{B}
-\overline{M}_{D_s} + \overline{M}_{D}}{1/2\,(1/M_D - 1/M_B)}
+
O\left(\frac{\Lambda_{QCD}^3}{m_Q}\right)\simeq - 0.06\pm 0.02\,
\mbox{GeV}^2\ .
\label{eq:deltal1def}\eeq
where the bar over the meson masses indicates ``spin average'', e.g.
$\overline{M}_B = (M_B + 3 M_{B^*})/4$, and $m_Q$ denotes the mass of the
$c$- or $b$-quark.
This value obtained from spectroscopy should be treated with some caution,
since its determination involves cancellations of much larger quantities,
and it is possible that higher-order terms affect the result. 
Similarly one can compute $\lambda_1(\Lambda_b)-\lambda_1(B_d)$, by 
extrapolating both the baryonic and mesonic kinetic energies to the 
chiral limit, and taking the differerence (or the other way round).
\section{Numerical calculation of $\lambda_2$} 
\label{numerical2}
The numerical calculation of $\lambda^{{\rm bare}}_2$ proceeds along
the same lines as the calculation of $\lambda^{{\rm bare}}_1$, so that
we will not repeat all the details given in subsection
\ref{sub:lambda1}. Also in this case, in order to compute the matrix
element of the chromo-magnetic operator, we used either
$R_{\vec{\sigma}\cdot \vec{G}}(t^\prime, t)$ or $R_{\vec{\sigma}\cdot
  \vec{G}}(t)=\sum_{t^\prime=0}^{t} R_{\vec{\sigma}\cdot
  \vec{G}}(t^\prime, t)$.  The UKQCD collaboration \cite{ukqcdl2} also
used the quantity $R_{\vec{\sigma}\cdot \vec{G}}(t)$. In this
case we make a linear fit of the form
\be R_{\vec{\sigma}\cdot \vec{G}}(t)=\alpha_2 + 3\, \lambda^{{\rm bare}}_2 t 
\, .\ee
\begin{figure} \vspace{9pt}
\begin{center}\setlength{\unitlength}{1mm} \begin{picture}(160,100)
\put(0,-58){\includegraphics{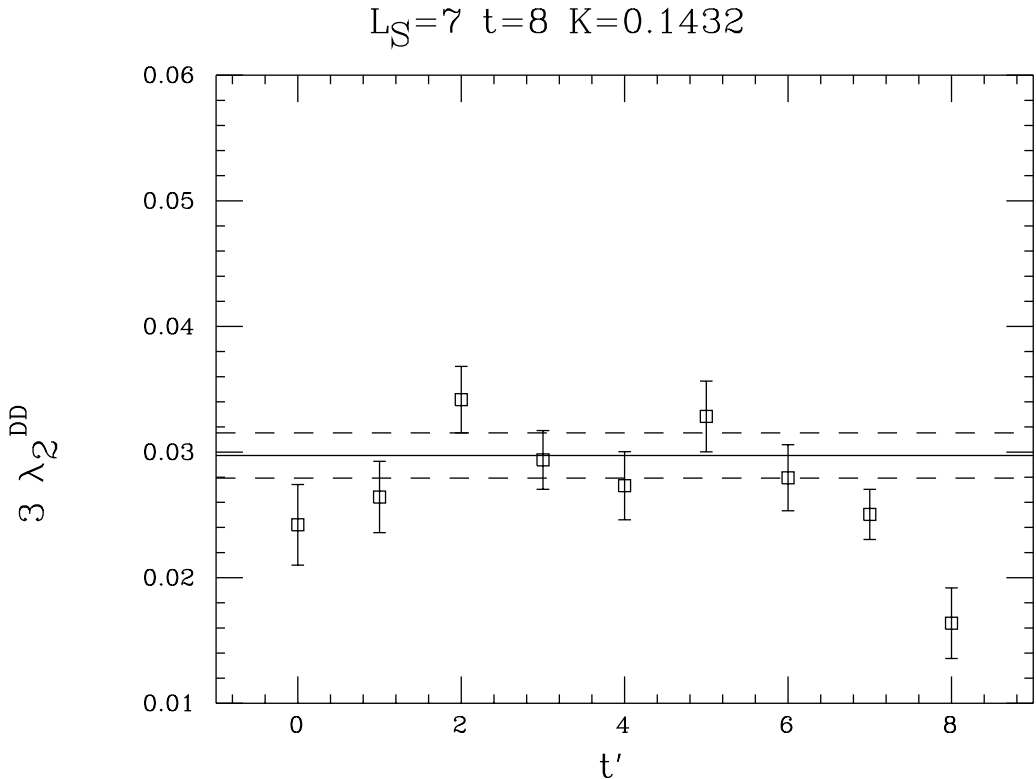}} \end{picture} \end{center}
\caption{\it{The ratio
$3 \, \lambda_2^{DD}= R_{\vec{\sigma}\cdot \vec{G}}(t^\prime, t)$, for $L_S=7$,
as a function of $t^\prime$,  at $t=8$.}}
\label{fig:l2pl}
\end{figure}
\begin{figure} \vspace{9pt}
\begin{center}\setlength{\unitlength}{1mm} \begin{picture}(160,100)
\put(0,-58){\includegraphics{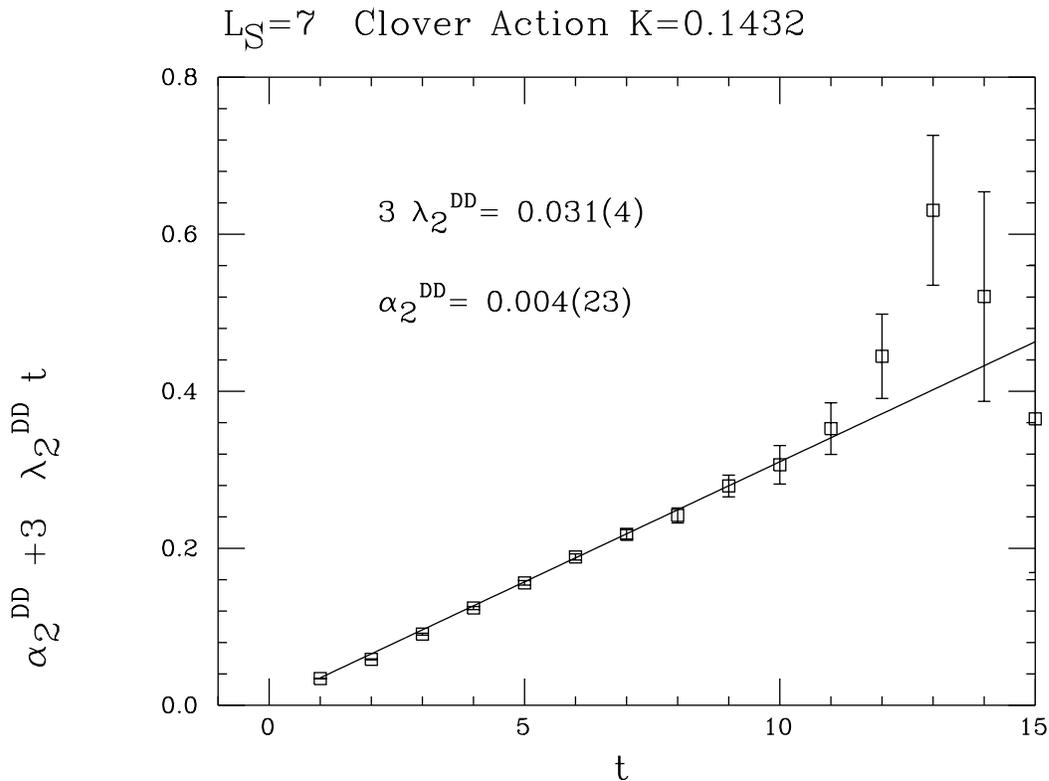}} \end{picture} \end{center}
\caption{\it{The ratio
$\alpha_2 + 3 \, \lambda_2^{{\rm bare}}t= R_{\vec{\sigma}\cdot \vec{G}}(t)$, for
$L_S=7$, as a function of $t$.}}
\label{fig:l2li}
\end{figure}

\par We now present our results.  In fig.~\ref{fig:l2pl} we
show $3 \, \lambda_2^{DD}=R_{\vec{\sigma}\cdot \vec{G}}(t^\prime, t)$ as a
function  of $t^\prime$, at fixed $t=8$,
obtained using the double-smeared sources $J^{D}_{B}(x)$, for $L_S=7$,
 at the  value of the light-quark Wilson parameter $K=0.1432$. 
In fig.~\ref{fig:l2li}
 we also give $R_{\vec{\sigma}\cdot \vec{G}}(t)$ as a function
of $t$, for the same sources, at the same value of the light-quark mass.
Notice that the large statistics has allowed us to perform the fit
up to rather large values of $t$ (as large as $t = 12$), 
thus reducing possible systematic uncertainties coming from 
higher excitations. For comparison the UKQCD
  collaboration fit $R_{\vec{\sigma}\cdot \vec{G}}(t)$ in the
  interval $ 2 \le t \le 5$. As in the case of
$\lambda^{{\rm bare}}_1$, we use the result obtained for a  smearing size $L_S=7$ with
double-smeared sources to determine the central value of $\lambda^{{\rm bare}}_2$, and
take into account the variations of the result with the smearing size and
time intervals to estimate the error. The main results are given in table
\ref{tab:l2res}. The value of $\lambda_2^{{\rm bare}}$ is consistent,
within the statistical errors, with $\lambda_2^{{\rm bare}}$ being independent
of the light-quark mass.
\begin{table} \centering
\begin{tabular}{||c|c|c||}
\hline
\hline
\multicolumn{3}{||c||}{$3 \, \lambda_2^{{\rm bare}}$ from 
$R_{\vec{\sigma}\cdot \vec{G}}(t^\prime,t)$
and $R_{\vec{\sigma}\cdot \vec{G}}(t)$ for $L_S=7$}\\
\hline \hline
$K$&
\multicolumn{1}{c|}{$t=8$, $t^\prime=3$--$5$}&
\multicolumn{1}{c|}{Fit with  $6 \le t \le 12$} \\
\hline \hline
0.1425& 0.0297(16)&0.031(3) \\
0.1432& 0.0297(18)&0.031(4) \\
0.1440& 0.030(2)&0.030(4) \\ \hline
0.145431&0.030(3) &0.029(5)\\
\hline\hline
\end{tabular}
\caption{\it{Values of $3 \, \lambda_2^{{\rm bare}}$ obtained as explained 
in the text.}}
\label{tab:l2res}
\end{table}
\par
In the case of $\lambda_2^{{\rm bare}}$ we observe a slightly larger
dependence of the results on the smearing size. We estimate the error
due to the imperfect isolation of the lightest state to be $0.005$.
Thus we quote \be 3 \, \lambda_2^{{\rm bare}}= 0.030 \pm 0.003 \pm
0.005 \, .\ee In order to obtain the matrix element of the
renormalized operator, we have evaluated the constant $Z_{\vec{\sigma}\cdot
  \vec{G}}$ using the one-loop perturbative calculation of
ref.~\cite{fh} and the same boosted coupling as in subsection
\ref{kinener}. This gives \be Z_{\vec{\sigma}\cdot \vec{G}}=1.85 \,
,\ee a very large correction that suggests that higher-order
contributions can be important (even with the standard definition of
the lattice strong coupling constant the one-loop correction,
corresponding to a value of $Z_{\vec{\sigma}\cdot \vec{G}}=1.51$, is
rather large).
\par 
Using the same calibration of the lattice spacing as before, we get \be
\lambda_2= 0.070 \pm 0.015 \quad \mbox{GeV}^2 \, , \ee where the error
combines in quadrature the statistical error, the systematic errors
discussed before, and the error from the calibration of the lattice
spacing. Our result, which corresponds to a mass splitting
$M^2_{B^*}-M^2_B= 4 \lambda_2= 0.280 \pm 0.060 $ GeV$^2$, although in
agreement with the calculation of refs.~\cite{ukqcdl2,bocl2}, is about
a factor of 2  smaller than the experimental result.  One may wonder
whether the discrepancy originates from the fact that the one-loop
perturbative corrections needed to obtain the matrix element of the
renormalized operator are so large~\footnote{ A non-perturbative
  calculation of $Z_{\vec{\sigma}\cdot \vec{G}}$ using the method
  proposed in ref.~\cite{NP} would help to clarify the situation.}, or
whether it is due to the quenched approximation, or to a physical
reason. Notice, however, that QCD sum rule determinations of the
splitting have given results that are in agreement with the
experimental value \cite{neubertl1,qcdsr}.  We believe, however, that
the possibility that our result for $\lambda_2$ is correct, and that
higher-order corrections in $1/m_Q$ are still important for the
$B^*$--$B$ mass difference, remains open. 
Using Wilson fermions for propagating quarks, one also
obtains hyperfine splittings which are smaller than the experimental
numbers. This is a different problem, however, which is 
related to the presence of a spurious chromo-magnetic term of $O(a)$
present in the lattice action.

\par From table~\ref{tab:l2res} we observe a very mild dependence of
$\lambda_2$ on the mass of the light quark, which would make $M_{B_s^*}^2
-M_{B_s}^2$ slightly larger than $M_{B^*}^2 -M_{B}^2$.

\section{Conclusions}
\label{conclu}
We have presented the results of a high-statistics lattice calculation
of the kinetic energy $\lambda_1$ and of the matrix element  
$\lambda_2$ of the
chromo-magnetic operator of the heavy quark in a $B$-meson in the HQET.
\par The results for $\lambda_1$   significantly improve a previous
calculation presented in ref.~\cite{cgms}. Our best estimate is
$\lambda_1 = 0.09\pm 0.14$~GeV$^2$.
In order to make a meaningful  comparison of 
this result with other theoretical determinations of the same
quantity, a perturbative calculation of the
relation between the different definitions of the renormalized
kinetic energy operator is needed, at a sufficient degree of accuracy 
\cite{ms2}. This calculation is missing to date, since both on the 
lattice and in the continuum only one-loop results are known.

\par For $\lambda_2$ we obtain the value $0.070 \pm 0.015$~GeV$^2$, which
corresponds to $M_{B^*}^2-M_B^2 = 0.280 \pm 0.060$~GeV$^2$, which is
about half of the experimental number $M_{B^*}^2-M_B^2 =
0.485$~GeV$^2$. This discrepancy, which is common to all
lattice results, could have several reasons, e.g. the
renormalization of the relevant operator, the quenched approximation,
or higher-order terms in the heavy-quark expansion, and needs to be
clarified.

\section*{Acknowledgements} 
We thank J.~Flynn and M.~Neubert for helpful discussions. 
We acknowledge the partial support by the EC
contract CHRX-CT92-0051. G.M. acknowledges the partial support by
M.U.R.S.T.  V.G.~wishes to thank the Istituto di Fisica ``G. Marconi''
of the Universit\`a di Roma ``La Sapienza'' for its hospitality,
and acknowledges partial support by CICYT under grant number 
AEN-96/1718.    
C.T.S. acknowledges the Particle Physics and Astronomy Research
Council for its support through the award of a Senior Fellowship.

\end{document}